\documentclass[journal]{IEEEtran}

\newcommand{\set}[1]{\mathcal{#1}}
\newcommand{\op}[1]{\mathrm{#1}}

\newcommand{\ve}[1]{\ensuremath{\mathbf{#1}}}

\usepackage{amsopn}
\DeclareMathOperator*{\argmin}{argmin}

\newcommand{\Gr}{\set{G}_\op{r}} % road network graph
\newcommand{\Nr}{\set{N}_\op{r}} % nodes road network graph
\newcommand{\Er}{\set{E}_\op{r}} % edges road network graph
\newcommand{\WGr}{L} % link lengths road network graph

\newcommand{\Assig}{\set{A}} % Assignment
\newcommand{\SPos}{\set{P}^\op{S}} % start position
\newcommand{\DPos}{\set{P}^\op{D}} % destination position
\newcommand{\x}{x} % continuous position
\newcommand{\xfun}{\xi} % continuous position trajectory (function)
\newcommand{\xfuni}[1]{\xfun_{#1}} % continuous position trajectory (function) of assignment i
\newcommand{\xS}{x^\op{S}} % start continuous position
\newcommand{\xD}{x^\op{D}} % destination continuous position
\newcommand{\e}{e} % edge
\newcommand{\efun}{\epsilon} % edge trajectory (function)
\newcommand{\efuni}[1]{\efun_{#1}} % edge trajectory (function) of assignment i
\newcommand{\eS}{e^\op{S}} % start edge
\newcommand{\eD}{e^\op{D}} % destination edge
\newcommand{\route}{\ve{e}} % route as sequence of edges
\newcommand{\routei}[1]{\route_{#1}} % route as sequence of edges
\newcommand{\en}[1]{\route[#1]} % nth edge in the route
\newcommand{\routeN}{N_{\op{e}}} % number of elements in route

\newcommand{\tS}{t^\op{S}} % start time
\newcommand{\tSi}[1]{\tS_{#1}} % start time of assignment i
\newcommand{\tD}{t^\op{D}} % arrival deadline
\newcommand{\tDi}[1]{\tD_{#1}} % arrival deadline of assignment i
\newcommand{\tA}{t^\op{A}} % actual arrival time
\newcommand{\tAi}[1]{t_{#1}^\op{A}} % actual arrival time of assignment i
\newcommand{\tM}{t^\op{M}} % merge time
\newcommand{\tMi}[1]{t_{#1}^\op{M}} % merge time of assignment i
\newcommand{\tSp}{t^\op{Sp}} % split time
\newcommand{\tSpi}[1]{t_{#1}^\op{Sp}} % split time of assignment i

\newcommand{\vehplan}{\set{P}} % vehicle plan
\newcommand{\vehplani}[1]{\set{P}_{#1}} % vehicle plan
\newcommand{\Dist}{D} % distance from start to goal, lenght of the route

\newcommand{\pind}{p} % platooning indicator
\newcommand{\pfun}{\pi} % platooning indicator profile (function)
\newcommand{\pfuni}[1]{\pfun_{#1}} % platooning indicator profile (function)
\newcommand{\pfunij}[2]{\pfun_{#1,#2}} % adapted platooning indicator profile i to j (a function)

\newcommand{\vt}{v} % speed
\newcommand{\vfun}{\phi} % speed profile (a function)
\newcommand{\vfuni}[1]{\vfun_{#1}} % speed profile (a function)
\newcommand{\vfunij}[2]{\vfun_{#1,#2}} % adapted speed profile i to j (a function)
\newcommand{\vseq}{\ve{v}} % sequence of vehicle speeds
\newcommand{\vseqi}[1]{\ve{v}_{#1}} % sequence of vehicle speeds, assignment i
\newcommand{\vseqn}[1]{\vseq[#1]} % element in sequence of vehicle speeds
\newcommand{\vseqin}[2]{\vseq_{#1}[#2]} % element in sequence of vehicle speeds for assignment i
\newcommand{\vseqN}{N_{\op{v}}} % number of elements in the speed profile
\newcommand{\vseqNi}[1]{N_{\op{v},#1}} % number of elements in the speed profile, assignment i

\newcommand{\vmin}{v_\op{min}} % min speed
\newcommand{\vmax}{v_\op{max}} % max speed
\newcommand{\vcm}{v_\op{cm}} % lowest possible assignment speed
\newcommand{\vcd}{v_\op{cd}} % fuel optimal constant speed

\newcommand{\tseq}{\hat{\ve{t}}} % sequence of vehicle speeds
\newcommand{\tseqi}[1]{\tseq_{#1}} % sequence of vehicle speeds, assignment i
\newcommand{\tseqn}[1]{\tseq[#1]} % nth element in sequence of vehicle speeds
\newcommand{\tseqin}[2]{\tseq_{#1}[#2]} % nth element in sequence of vehicle speeds for assignment i

\newcommand{\Ftot}{F_\op{c}} % total fuel consumption of the plan
\newcommand{\Fvp}[2]{F(#1,#2)} % total fuel consumption of a trajectory
\newcommand{\f}{f} % fuel consumption per distance traveled as function of speed and platooning indicator
\newcommand{\fvp}[2]{f(#1,#2)} % fuel consumption per distance traveled as function of speed and platooning indicator with parameters
\newcommand{\fzero}{f_0} % regular fuel consumption per distance traveled as function of speed
\newcommand{\fzerov}[1]{f_0(#1)} % regular fuel consumption per distance traveled as function of speed with parameters
\newcommand{\fplat}{f_\op{p}} % regular fuel consumption per distance traveled as function of speed
\newcommand{\fplatv}[1]{f_\op{p}(#1)} % regular fuel consumption per distance traveled as function of speed with parameters
\newcommand{\fce}{f_\op{ce}} % central objective clustering

\newcommand{\Gc}{\set{G}_\op{c}} % coordination graph
\newcommand{\Nc}{\set{N}_\op{c}} % nodes coordination graph, assignment set
\newcommand{\Ec}{\set{E}_\op{c}} % edges coordination graph
\newcommand{\WGc}{\Delta F} % weights coordination graph
\newcommand{\WGcij}[2]{\WGc(#1,#2)} % weights coordination graph from i to j

\newcommand{\Nin}[1]{\set{N}_{#1}^{\op{i}}} % set of in neighbors of node i
\newcommand{\Nout}[1]{\set{N}_{#1}^{\op{o}}} % set of in neighbors of node i

\newcommand{\Nl}{\set{N}_\op{l}} % leader set

\newcommand{\U}{\set{U}} % universe
\newcommand{\FS}{\set{S}_{\op{u}}} % family of subsets
\newcommand{\SpS}{\set{S}} % specific subset
\newcommand{\sps}{\set{S}} % specific subset iterrator
\newcommand{\NO}{N_1} % top node
\newcommand{\NA}{\set{N}_2} % subset level nodes
\newcommand{\NB}{\set{N}_3} % universe level nodes
\newcommand{\mA}{\mu_2} % subset level nodes map
\newcommand{\mB}{\mu_3} % universe level nodes map

\newcommand{\deltau}{\Delta u} % weights coordination graph from i to j
\newcommand{\deltauij}[2]{\deltau(#1,#2)} % weights coordination graph from i to j
\newcommand{\n}{n} % the node that is changed

\newcommand{\nl}{n_\op{l}} % leader of the cluster in joint speed profile optimization
\newcommand{\Nfl}[1]{\set{N}_{\op{fl},#1}} % set of followers of node i
\newcommand{\Ng}{\set{N}_{\op{g}}} % coordination group

\newcommand{\tcon}{\ve{t}} % merge and split times sequence
\newcommand{\tconn}[1]{\tcon[#1]} % merge and split times sequence, nth element

\newcommand{\Wcon}{\ve{W}} % distance sequence
\newcommand{\Wconi}[1]{\Wcon_{#1}} % distance sequence for assignment i
\newcommand{\Wconin}[2]{\Wcon_{#1}[#2]} % distance sequence for assignment i, element n

\newcommand{\indMi}[1]{i_{#1}^\op{M}} % index of the first element in Wcon of the leader the follower and leader have in common
\newcommand{\indSpi}[1]{i_{#1}^\op{Sp}} % index of the last element in Wcon of the leader the follower and leader have in common

\newcommand{\pcon}{\ve{p}} % platooning sequence
\newcommand{\pconi}[1]{\pcon_{#1}} % platooning sequence for assignment i
\newcommand{\pconin}[2]{\pcon_{#1}[#2]} % platooning sequence for assignment i, element n

\newcommand{\Tcon}{\ve{T}} % platooning sequence for assignment i
\newcommand{\Tconi}[1]{\Tcon_{#1}} % platooning sequence for assignment i
\newcommand{\Tconin}[2]{\Tcon_{#1}[#2]} % platooning sequence for assignment i, element n

\usepackage{amsthm}

\newtheorem{prop}{Proposition}

\theoremstyle{definition}

\newtheorem{defi}{Definition}
\newtheorem{problem}{Problem}

\usepackage[cmex10]{amsmath}
\usepackage{amssymb}
\usepackage{mathtools}
\usepackage[utf8]{inputenc} 

\usepackage{algorithmic}

\usepackage{color}

\usepackage{url}

\usepackage{cite}

\ifCLASSINFOpdf
  \usepackage[pdftex]{graphicx}
  \graphicspath{{plots/}}
\else
\fi

\usepackage{array}

\begin{document}

\title{Fuel-Efficient En Route Formation\\of Truck Platoons}

\author{Sebastian~van~de~Hoef,~\IEEEmembership{}
        Karl~H.~Johansson,~\IEEEmembership{Fellow, IEEE}
        and~Dimos~V.~Dimarogonas,~\IEEEmembership{Member, IEEE}% <-this % stops a space
\thanks{ACCESS Linnaeus Center and the School of Electrical Engineering, KTH Royal Institute of Technology, SE-100 44, Stockholm, Sweden (e-mail: {\tt \{shvdh,kallej,dimos\}@kth.se}).}}

\markboth{}%
{}

\maketitle

\begin{abstract}
 The problem of how to coordinate a large fleet of trucks with given itinerary to enable fuel-efficient platooning is considered. Platooning is a promising technology that enables trucks to save significant amounts of fuel by driving close together and thus reducing air drag. A setting is considered in which each truck in a fleet is provided with a start location, a destination, a departure time, and an arrival deadline from a higher planning level. Fuel-efficient plans should be computed. The plans consist of routes and speed profiles that allow trucks to arrive by their arrival deadlines. Hereby, trucks can meet on common parts of their routes and form platoons, resulting in decreased fuel consumption. 
 We formulate a combinatorial optimization problem that combines plans involving only two vehicles. We show that this problem is hard to solve for large problem instances. Hence a heuristic algorithm is proposed. The resulting plans are further optimized using convex optimization techniques. The method is evaluated with Monte Carlo simulations in a realistic setting. We demonstrate that the proposed algorithm can compute plans for thousands of trucks and that significant fuel savings can be achieved. 
\end{abstract}

\IEEEpeerreviewmaketitle

\section{Introduction}

\IEEEPARstart{P}{latooning} is foreseen to become a common element in intelligent transportation systems. The term refers to a group of vehicles forming a road train without any physical coupling between them. A short inter-vehicle distance is maintained by automatic control and vehicle-to-vehicle communication. 
Platooning has received a lot of attention due to its potential to increase road throughput by reducing the inter-vehicle gaps. It can also help facilitate the \mbox{(semi-)}automatic operation of vehicles~\cite{path_overview_conference}.
This paper focuses on the potential of platooning in reducing fuel consumption.
Similar to what racing cyclists exploit, the follower vehicles, and to a lesser degree the lead vehicle, experience a reduction in air drag, which translates into reduced fuel consumption \cite{Bonnet2000, lammert_fuel_consumption, ITS_overview}. Advances in wireless communication, satellite-based position, and advanced driver support systems have made the wide deployment of platooning systems feasible and have attracted the attention of major truck manufacturers. Increased fuel costs and environmental awareness make the implementation of such systems likely in the near future.

Using platooning to reduce fuel consumption for a large number of trucks leads to a challenging coordination problem. 
Consider two trucks that travel between the same two regions but from different locations and at approximately the same time. Then the trucks can adjust their speeds slightly at the beginning of their journeys, form a platoon at the start of the common part of their route and thus save fuel during part of their trips. This approach might involve one of the trucks having to drive slightly faster before the two merge, which increases air-drag and consequently fuel consumption during the initial phase. One truck might instead slow down to let the other truck catch up but then travel at an increased speed later on to arrive at its destination on time. If many trucks are involved, it is not straightforward to compute an optimal plan for all trucks. 

The main contribution of the paper is to derive an efficient and scalable method to coordinate platooning of a large number of trucks in a fuel-efficient way explicitly considering the effect of speed and platooning on the fuel consumption. The core novelty is that the computation of platoon plans is done in three computationally and mathematically tractable stages. The first stage involves the computation of platoon plans only taking into account two vehicles at a time. The second stage selects one such plan for each vehicle. Since the problem to be solved in the second stage is shown to be NP-hard, an iterative algorithm to compute heuristic solutions is proposed. In the third stage, the resulting plans are optimized further using convex optimization techniques. The potential of the method is demonstrated using Monte Carlo simulations. The simulations demonstrate that the method is able to significantly reduce the fuel consumption of a fleet of vehicles as well as to handle a realistic number of transport assignments for the example of Sweden.

The efficient operation of transport systems is a widely studied field due to its large economic and environmental impact. Planning conducted by transportation operators ranges all the way from strategic over tactical to operational planning~\cite{FreightPlanningSurvey}. The planning on the latter stage typically happens on the level of departure and arrival times which is the input to the problem considered in this paper. Research on eco-routing aims to reduce fuel consumption by appropriate choice of route and travel speeds for individual vehicles~\cite{eco_routing_journal}. Furthermore, operators of road infrastructure use variable speed limits, ramp metering, variable route recommendations, etc. to improve safety and efficiency of the road transportation system \cite{road_traffic_control}. 

Various aspects of platoon coordination have been considered in the literature.
In \cite{Larsson_Platoon_Complexity} the authors formulate a mixed integer linear programming problem, without considering the speed dependency of fuel consumption, and prove that the problem is NP-hard. In \cite{kuoyun_catchup} the authors consider a simple catch-up coordination scheme and evaluate it on real fleet data. In \cite{jeff_kuo_yun_distributed_controller} local controllers for coordinating the formation of platoons are proposed. In \cite{datamining_platooning} the authors use data-mining to identify economic platoons based on various criteria. Unlike this paper the method presented in \cite{datamining_platooning} allows that trucks wait for other trucks to form the platoon. Preliminary material used in this paper has been presented in \cite{ACCpaper}, \cite{ITSCpaper}. 

One of the key elements to make the problem tractable for realistic numbers of trucks, is to select a subset of vehicles called coordination leaders to which other vehicles adapt. The way we select this subset of vehicles is inspired by a clustering algorithm called partitioning around medoids \cite{pam_book}.
Clustering is present in a variety of different contexts. A large body of research focuses on clustering methods as an analysis tool to structure, understand, and classify large data sets \cite{clustering_book1, clustering_overview_paper}. Examples include the clustering of graphs in the scope of community detection \cite{blondelcommunity, community_detection_survey, community_detection_survey2}. 
Closely related to our problem of choosing coordination leaders is leader election, where a group of agents has to jointly determine a leader \cite{bully_alg, electing_good_leaders}. The approach developed in this paper can be seen as local leader election, where pairwise fuel savings are interpreted as preferences for trucks being coordination leaders.

The paper is organized as follows. We start in Section~\ref{sec:problem_formulation} by formulating the coordination problem and introduce the structure of the proposed solution in Section~\ref{sec:platoon_coordinator}. In Section~\ref{sec:clustering} the problem is broken down to a purely combinatorial problem of selecting coordination leaders and the problem is shown to be NP-hard. In Section~\ref{sec:clustering_alg} an iterative algorithm is developed to find heuristic solutions to the combinatorial problem.  
Section~\ref{sec:joint_speed-profile_optimization} discusses how to jointly optimize the plans when the trucks are constrained to the platoons as proposed by the algorithm from Section~\ref{sec:clustering_alg}. In Section~\ref{sec:simulations} the method is demonstrated by Monte Carlo simulations in a realistic scenario with up to 5\,000 trucks on the Swedish road network.

\section{Problem Formulation}
\label{sec:problem_formulation}

In this section we formulate the problem and introduce the notation.
We have an index set $\Nc$ of finitely many transport assignments, each tied to a specific truck. A transport assignment $\Assig = (\SPos,\DPos,\tS,\tD)$ consists of a start position $\SPos$, a destination $\DPos$, a start time $\tS$, and an arrival deadline $\tD$. We model the road network as a directed graph $\Gr = (\Nr, \Er)$ with nodes $\Nr$ and edges $\Er$. Nodes correspond to intersections or endpoints in the road network and edges correspond to road segments connecting these intersections. The function $\WGr: \Er \rightarrow \mathbb{R}^+$ maps each edge in $\Er$ to the length of the corresponding road segment. A vehicle position is a pair $(\e,\x) \in \Er \times [0, \WGr(\e)]$ where $\e$ indicates the current road segment and $\x$ how far the vehicle has traveled along that segment.

The goal is to compute fuel-efficient plans for the trucks that ensure arrival before each trucks' individual deadline. Each plan includes a route in the road network from start to destination and encodes a piecewise constant speed trajectory. The speed is constrained to a range of feasible speeds $[\vmin, \vmax]$, which is supposed to be the same for all vehicles and road segments.\footnote{The approach developed in this paper can be generalized in order to relax this assumption.} For the sake of this high-level planning, it is reasonable to assume that trucks change their speed instantaneously.
\begin{defi}[Vehicle Plan]\label{def:vehicle_plan}
  A vehicle plan $\vehplan = (\route,\vseq,\tseq)$ consists of a route $\route$, a speed sequence $\vseq$, and a time sequence $\tseq$. The route is a sequence of $\routeN$ edges in the road network: $\route = (\en{1}, \dots, \en{\routeN})$, $\en{i} \in \Er$. The speed sequence is a sequence of $\vseqN$ speeds $\vseq = (\vseqn{1}, \dots, \vseqn{\vseqN})$, where speeds are within the feasible speed range $0 < \vmin \leq \vseqn{i} \leq \vmax$. The time sequence $\tseq = (\tseqn{1}, \dots, \tseqn{\vseqN+1})$ defines when the speed changes. Speed $\vseqn{i}$ is selected from $\tseqn{i}$ until $\tseqn{i+1}$.  
\end{defi}
Note that changes in speed can happen, in principle, everywhere and not only at the beginning of route segments.
The symbols $\routeN$ and $\vseqN$ are introduced for notational convenience and their value may be different for different vehicle plans. 

We want to compute a vehicle plan for each truck. A valid vehicle plan brings the truck from its start position $\SPos = (\eS,\xS)$, where it is at time $\tS$, to its destination $\DPos= (\eD,\xD)$ before its deadline $\tD$. 

Vehicle plans are constrained by two conditions. The first condition requires the trip to start at the start time $\tseqn{1} = \tS$ and ends before the deadline $\tseqn{\vseqN+1} = \tA \leq \tD$. The second condition ensures that the truck arrives at its destination when the trip ends, i.e., the distance traveled is
\begin{equation*}
  \Dist \coloneqq \sum\limits_{i = 1}^{\routeN-1} \WGr(\en{i}) + \xD - \xS
  = \sum\limits_{i = 1}^{\vseqN} \vseqn{i} (\tseqn{i+1}-\tseqn{i}). \label{eq:arrival_cond}
\end{equation*}
We introduce the notion of trajectories as functions of continuous time.
A vehicle trajectory consists of an edge trajectory $\efun$ and a linear position trajectory $\xfun$. The edge trajectory for $t \in [\tS, \tA)$ is given by $\efun(t) = \en{j}$ where $j$ depends on $t$ and is the largest integer that satisfies $\sum\limits_{i = 1}^{j-1} \WGr(\en{i}) - \xS < \int\limits_{\tS}^{t} \vfun(\tau)\op{d}\tau$, and where the speed trajectory $\vfun(t) = \vseqn{i}$ for $t \in [\tseqn{i}, \tseqn{i+1})$, $i \in \{1, \dots, \vseqN\}$. The linear position, i.e., the second element of the position, at time $t$ is given by $\xfun(t) =\int\limits_{\tS}^{t} \vfun(\tau) \op{d}\tau - \sum\limits_{i = 1}^{j-1} \WGr(\en{i}) + \xS$. 

When trucks platoon, their positions coincide in our model and hence we neglect the physical dimension of the trucks. Each platoon consists of a platoon leader and a number of platoon followers. We introduce the platoon trajectory $\pfuni{n}: [\tSi{n}, \tAi{n}) \rightarrow \{0,1\}$ for truck $n \in \Nc$. A platoon trajectory equals $1$ when truck $n$ is a platoon follower and $0$ when it is a platoon leader or traveling alone. Thus $\pfuni{n}(t) = 1$ implies that there is another truck $m \in \Nc$ with $m \neq n$ and $(\efuni{n}(t),\xfuni{n}(t)) = (\efuni{m}(t), \xfuni{m}(t))$.

We model the fuel consumption per distance traveled as a function of the speed and of whether the truck is a platoon follower or not. A platoon leader is assumed to have the same fuel consumption as a truck that travels alone while a platoon follower has a reduced fuel consumption. Hereby, we neglect the relatively small reduction in fuel consumption when a truck is a platoon leader compared to traveling by itself.
We denote the fuel consumption per distance traveled as $\f: [\vmin, \vmax] \times \{0,1\} \rightarrow \mathbb{R}^+$ where
\begin{equation}
 \fvp{\vt}{\pind} = \left\{
 \begin{array}{ll}
  \fzerov{\vt} & \text{if } \pind = 0\\
  \fplatv{\vt} & \text{if } \pind = 1.
 \end{array}
 \right. \label{eq:fuel_consumption_combined}
\end{equation}
The function $\fzero$ models the fuel consumption when the truck is a platoon leader or when it travels solo, and $\fplat$ the fuel consumption when the truck is a platoon follower. These functions can either be derived from an analytical model or fitted to data \cite{fuel_model_review}. We purposely omit that fuel consumption depends on road and vehicle parameters in order to keep the presentation concise. All the presented results can be augmented to handle those additional parameters.

The problem that we want to solve is to find a vehicle plan for each vehicle, and we want to minimize the combined fuel consumption of these plans. The total fuel consumption $\Fvp{\vfuni{n}}{\pfuni{n}}$ associated with vehicle $n$'s plan is given by integrating the fuel consumption according to \eqref{eq:fuel_consumption_combined} over the duration of the trip
\begin{equation}
 \Fvp{\vfuni{n}}{\pfuni{n}} = \int\limits_{\tSi{n}}^{\tAi{n}} f(\vfuni{n}(t),\pfuni{n}(t)) \vfuni{n}(t) \op{d} t, \label{eq:fuel_consumption_integral}
\end{equation}
where $\vfuni{n}$ is the speed trajectory, $\pfuni{n}$ the platoon trajectory, $\tSi{n}$ the start time, and $\tAi{n}$ the arrival time of truck $n$.
The combined fuel consumption $\Ftot$ is given by
\begin{equation}
 \Ftot = \sum\limits_{n \in \Nc} \Fvp{\vfuni{n}}{\pfuni{n}}. \label{eq:overall_fuel_consumption}
\end{equation}
Our primary goal is to compute vehicle plans that minimize $\Ftot$. 

\section{Platoon Coordinator}
\label{sec:platoon_coordinator}

\begin{figure}[t]
\begin{center}
 \includegraphics[width=.9\columnwidth]{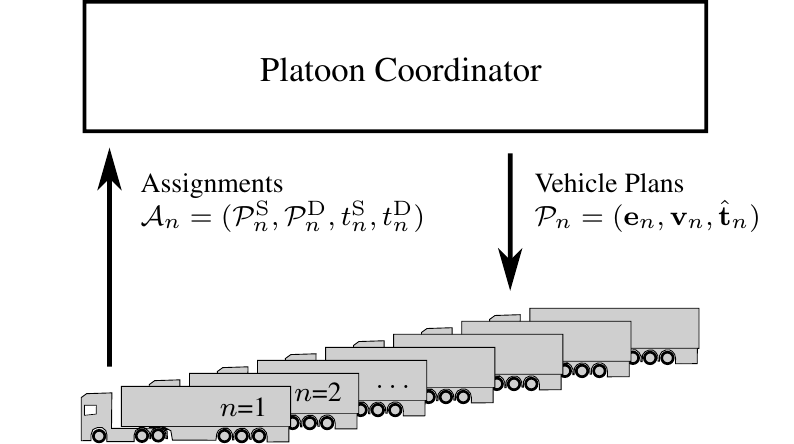}
 \caption{Schematic of the platoon coordinator. Trucks provide their assignment data and the platoon coordinator computes fuel efficient vehicle plans.}
\label{fig:system_overview}
\end{center}
\end{figure}

Consider the centralized platoon coordinator in Fig.~\ref{fig:system_overview}. Trucks connect to the coordinator via vehicle-to-infrastructure communication and share their assignment data. The coordinator then computes fuel-efficient vehicle plans for the trucks. These plans are sent to the trucks and executed. This process is repeated whenever there is updated information, such as deviations from the plans and new assignments. The current vehicle position is then the new start position of an assignment that is already being executed.

The computation of the vehicle plans happens in four stages:
\begin{enumerate}
 \item Computation of the routes $\routei{n}$, $n \in \Nc$: routes are calculated using an algorithm for route calculation in road networks.
 \item Computation of pairwise vehicle plans: many plans involving two vehicles are computed. The fuel savings of these plans are recorded as the coordination graph $\Gc$ introduced in the following. 
 \item Selection of pairwise plans: a consistent subset of the plans computed in the previous stage is combined by selecting a subset $\Nl \subset \Nc$, so-called called coordination leaders.
 \item Joint speed-profile optimization: the selected pairwise plans are jointly optimized for lower fuel consumption while platoon partners as well as merge and split locations are kept as computed in step 3).
\end{enumerate}
Stage 1) computes the routes $\routei{n}$, $n \in \Nc$ and stages 2)-4) compute the speed sequences $\vseqi{n}$ and time sequences $\tseqi{n}$ for $n \in \Nc$ making use of the ability of the trucks to form platoons in order to achieve low fuel consumption. Algorithms for route calculation in road networks are well developed \cite{routing_overview, eco_routing_journal} and not further discussed in this paper. We discuss stages 2)-4) in the following sections.

\section{Selecting Pairwise Vehicle Plans}
\label{sec:clustering}

In this section, we formulate a combinatorial optimization problem corresponding to the second and third computation stage introduced in Section~\ref{sec:platoon_coordinator}. 
The problem is proven to be NP-hard which motivates the heuristic algorithm developed in Section~\ref{sec:clustering_alg}. 

To begin with, we need to be able to compute what we call a default plan. This is a valid vehicle plan according to Definition~\ref{def:vehicle_plan} with either the lowest possible or most fuel optimal constant speed. 

\begin{defi}[Default Plan]
The default plan is a vehicle plan $\vehplan = (\route,\vseq,\tseq)$ with speed sequence $\vseq = (\vcd)$ and time sequence $\tseq = (\tS,  \Dist/\vcd)$.
The most fuel optimal speed without platooning $\vcd$ is computed as
\begin{align*}
 \vcd = \argmin\limits_{v \in (\vcm, \vmax]} \fzerov{\vt},
\end{align*}
where $\vcm$ is the lowest constant speed to arrive before the deadline:
\begin{align*}
 \vcm = \max \left(\vmin, \frac{\Dist}{\tD - \tS}\right).
\end{align*}
\end{defi}
An adapted plan, as introduced next, is such that the speed sequence $\vseqi{n}$ and time sequence $\tseqi{n}$ of a follower truck $n$ is adapted in a way that allows the follower to platoon during part of its journey with a leader $m$. The leader sticks to its default plan, which is important in order to be able to compose these plans. The plan is computed in a way that minimizes the fuel consumption of $n$.   
\begin{defi}[Adapted Plan] 
An adapted plan is a vehicle plan $\vehplani{n} = (\routei{n},\vseqi{n},\tseqi{n})$ adapted to vehicle plan $\vehplani{m} = (\routei{m},\vseqi{m},\tseqi{m})$, such that $(\efuni{n}(t),\xfuni{n}(t)) = (\efuni{m}(t),\xfuni{m}(t))$ for $t \in [\tseqin{n}{2}, \tseqin{n}{\vseqNi{n}})$.
\end{defi}

\begin{figure}[t]
\begin{center}
 \includegraphics[width=\columnwidth]{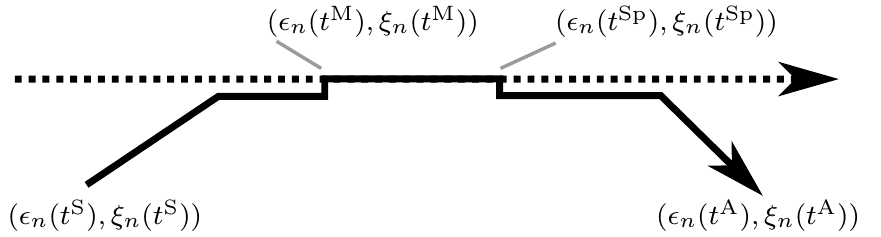}
 \caption{Overview of the relevant time instances of the adapted plan. The solid line illustrates the route of the adapted plan $n$ and the dashed line the one of the plan that it is adapted to and has index $m$. The parallel sections of the line indicate that the trucks share the route and the section where the lines are on top of each other indicates that the trucks platoon there.}
\label{fig:adapted_plan_illustration}
\end{center}
\end{figure}

We denote the merge time as $\tM = \tseqin{n}{2}$ and the split time as $\tSp = \tseqin{n}{\vseqNi{n}}$.
Truck $n$ becomes the platoon follower of truck $m$ at time $\tM$, stays platoon follower until $\tSp$, when the two trucks separate. This sequence of events occurs only once. Fig.~\ref{fig:adapted_plan_illustration} illustrates the adapted plan. We denote the speed trajectory $\vfun$ corresponding to the speed sequence $\vseq$ and the time sequence $\tseq$ of the adapted vehicle plan of truck $n$ adapted to truck $m$ as $\vfunij{n}{m}$. 

The fuel consumption of truck $n$ with its plan adapted to truck $m$ is modeled as in \eqref{eq:fuel_consumption_integral}. We denote the platoon trajectory of the adapted plan $\pfunij{n}{m}(t)$. We have that $\pfunij{n}{m}(t) = 1$ for $t \in [\tM, \tSp)$ and $\pfunij{n}{m}(t) = 0$ for $t \in [\tS, \tM) \cup [\tSp, \tA)$. The fuel consumption of $m$ is not altered by the fact that $n$ and $m$ platoon, since $m$'s speed trajectory does not change and since $m$ takes the role of a platoon leader. The reduction in fuel consumption that results from $n$ implementing the adapted plan and not its default plan is $\WGcij{n}{m} = \Fvp{\vfuni{n}}{\pfuni{n}} - \Fvp{\vfunij{n}{m}}{\pfunij{n}{m}}$ where $\pfuni{n} \equiv 0$, which is positive if $n$ adapting to $m$ saves fuel. If no plan that is adapted to $m$ exists for $n$, we define $\WGcij{n}{m} = 0$. There might exist no adapted plan because the routes do not overlap or because the constraint on the maximum speed in conjunction with the arrival deadline makes it impossible for the trucks to form a platoon. 

We now compute $\WGc$ for all 2-permutations in $\Nc$. We are only interested in adapted plans that save fuel, i.e., for which $\WGc$ is positive. We can conveniently collect this information in a weighted graph that we call the coordination graph.
\begin{defi}[Coordination Graph]
 The coordination graph is a weighted directed graph $\Gc = (\Nc,\Ec,\WGc)$. Recall that the elements of $\Nc$ represent the trucks. $\Ec \subseteq \Nc \times \Nc$ is a set of edges, and $\WGc: \Ec \rightarrow \mathbb{R}^+$ are edge weights, such that there is an edge $(n,m) \in \Ec$, if the adapted plan of $n$ to $m$ saves fuel compared to $i$'s default plan, i.e., $\Ec =  \{(i,j) \in \Nc \times \Nc: \WGcij{i}{j} > 0, i \neq j\}$.
\end{defi}
Furthermore, we introduce the set of in-neighbors of a node $n \in \Nc$ as $\Nin{n} = \{i \in \Nc: (i,n) \in \Ec\}$ and the set of out-neighbors $n$ as $\Nout{n} = \{i \in \Nc: (n,i) \in \Ec\}$. We define the the maximum over an empty set to be zero, i.e., $\max\limits_{i \in \emptyset}(\cdot) = 0$. 

With these definitions, we are ready to formulate the problem of finding a fuel optimal set of coordination leaders $\Nl$.
\begin{problem}\label{prob:clustering}
Given as input a coordination graph $\Gc = (\Nc,\Ec,\WGc)$ find a subset $\Nl \subset \Nc$ of nodes that maximizes
\begin{equation}
 \fce(\Nl) = \sum\limits_{i \in \Nc \setminus \Nl} \max_{j \in \Nout{i} \cap \Nl}  \WGcij{i}{j}.  \label{eq:clustering_problem}
\end{equation}
\end{problem}
The coordination leaders select their default plans. The remaining assignments, called coordination followers, select their plans adapted to the coordination leader that yields the largest fuel savings $\WGcij{n}{m}$. Since the selection of adapted plans does not alter the speed trajectories of the coordination leaders, several coordination followers can select the same coordination leader without affecting the fuel savings that result from this adaptation, potentially resulting in platoons of more than two vehicles. The objective function $\fce(\Nl)$ equals the sum of all these fuel savings. 
If $(n,m) \in \Ec$ with $n \in \Nc \setminus \Nl$ and $m = \arg \max\limits_{m \in \Nout{n} \cap \Nl}\WGcij{n}{m}$, we say that $n$ is the coordination follower of $m$ and $m$ is the coordination leader of $n$. If $m$ has no out-neighbor in $\Nl$, then $\max_{m \in (\Nout{n} \cap \Nl)} \WGcij{n}{m}  = \max_{m \in \emptyset}  \WGcij{n}{m}  = 0$. 

Problem~\ref{prob:clustering} is a combinatorial optimization problem. We can compute an optimal solution in finite time by using exhaustive search or a branch and bound technique. However, the computational complexity of such an exact computation might be too high. In fact, we show in the following that this problem is an NP-hard problem. This a strong indicator that searching for an algorithm that computes solutions for every coordination graph and scales well in the size of the coordination graph is futile \cite{intro_to_algorithms}.

\begin{prop}\label{prop:NP}
 Problem~\ref{prob:clustering} is NP-hard, where the size of the input $\Gc$ is measured as $|\Nc| + |\Ec|$.
\end{prop}
The proof can be found in the appendix.

One disadvantage of the approach presented in this section is that each truck can only join one platoon. This can however be somewhat mitigated by frequent re-planning. For instance, at some later point in time, it might turn out more beneficial for a truck to leave its current platoon and join another one.

\section{Iterative Selection of Coordination Leaders}
\label{sec:clustering_alg}

In this section we present an algorithm that computes heuristic solutions to Problem~\ref{prob:clustering}. Motivated by the result that Problem~\ref{prob:clustering} is NP-hard, we apply an iterative strategy that converges to a local maximum.

\begin{figure}[t]
\begin{algorithmic}
\REQUIRE $\Gc$
\ENSURE $\Nl$

\STATE $\Nl \leftarrow \emptyset$
\WHILE{$\{\bar{n} \in \Nc: \deltauij{\bar{n}}{\Nl} > 0\} \neq \emptyset$}
  \STATE Select $\n \in \{\bar{n} \in \Nc: \deltauij{\bar{n}}{\Nl} > 0\}$
    \IF {$\n \in \Nl$}
      \STATE $\Nl \leftarrow \Nl \setminus \{\n\}$
    \ELSE
      \STATE $\Nl \leftarrow \Nl \cup \{\n\}$
    \ENDIF
\ENDWHILE
\end{algorithmic}
\caption{The clustering algorithm, which is an iterative algorithm to compute the set of coordination leaders $\Nl$.}\label{alg:general_alg}
\end{figure}

Consider the algorithm in Fig.~\ref{alg:general_alg}, which we call the clustering algorithm. The input is a coordination graph $\Gc$ and the output is a set of coordination leaders $\Nl$. Initially $\Nl$ is an empty set. In each iteration, a node $\n \in \Nc$ is selected for which the objective function $\fce$ is increased if it is added to $\Nl$ or removed from $\Nl$, and $\Nl$ is updated accordingly. The difference in $\fce$ when adding or removing a node in $\Nc$ to or from the set of coordination leaders $\Nl$ is given by a function $\deltau$. The algorithm iterates until no further increase of $\fce$ is possible. 

The function $\deltau$ that measures how much is gained from switching whether $\n$ belongs to $\Nl$ is defined as follows:
\begin{align}
 \deltauij{\n}{\Nl} = \left\{ \begin{array}{ll}
                   \fce(\Nl \setminus \{\n \}) - \fce(\Nl) & \text{ if } \n \in \Nl\\
                   \fce(\Nl \cup \{\n\}) - \fce(\Nl) & \text{ otherwise }.
                  \end{array} \right.\label{eq:delta_u_c}
\end{align}
If $\n \notin \Nl$, we get
\begin{align*}
 &\fce(\Nl \cup \{\n\}) - \fce(\Nl)=\\
 &\sum\limits_{i \in \Nin{n}\setminus \Nl} \left( 
 \max\limits_{j \in \Nout{i} \cap (\Nl \cup \{\n\})} \WGcij{i}{j}
 - \max\limits_{j \in \Nout{i} \cap \Nl} \WGcij{i}{j}
 \right)\\
 &- \max\limits_{i \in \Nout{\n} \cap \Nl} \WGcij{\n}{i}.
\end{align*}
The sum over $i$ covers nodes that can select $n$ as their new coordination leader. The last summand accounts for $n$ possibly not being a coordination follower any longer.

If $\n \in \Nl$, we get
\begin{align*}
 &\fce(\Nl \setminus \{\n\}) - \fce(\Nl) =\\
 &\sum\limits_{i \in \Nin{\n}\setminus \Nl} \left( 
 \max\limits_{j \in \Nout{i} \cap (\Nl \setminus \{\n\})} \WGcij{i}{j}
 - \max\limits_{j \in \Nout{i} \cap \Nl} \WGcij{i}{j}
 \right)\\
 &+ \max\limits_{i \in \Nout{\n} \cap (\Nl \setminus \{\n\})} \WGcij{n}{i}. % W_\op{c}(n,n) = 0
\end{align*}
The sum over $i$ covers nodes that can have $\n$ as their coordination leader before the change. The last summand accounts for $\n$ possibly becoming a coordination follower. 

In this paper, we consider two methods to select $\n$ from the set $\{\bar{n} \in \Nc: \deltauij{\bar{n}}{\Nl} > 0\}$.
The first method is to select $n$ in a greedy manner according to $\n = \arg \max\limits_{\bar{\n} \in \Nc} \deltauij{\bar{\n}}{\Nl}$. The second method is to choose $\n$ randomly with equal probability from the set $\{\bar{n} \in \Nc: \deltauij{\bar{n}}{\Nl} > 0\}$. 

The clustering algorithm is guaranteed to converge in finite time. This is due to the number of possible subsets of $\Nc$ being finite and thus the number of possible assignments of $\Nl$ is finite. In every iteration $\fce(\Nl)$ strictly increases which means that $\Nl$ changes in every iteration and the same assignment for $\Nl$ never reoccurs. So in the worst case the clustering algorithm iterates over all subsets of $\Nc$ before termination.

The clustering algorithm can be efficient. Note for instance that the function $\deltau$ can be computed based on the sub-graph induced by the one- and two-hop neighbors of $\n$ only. This means that the average complexity of computing $\Delta u$ is a function of the average node degree and not of the number of nodes in the coordination graph. Furthermore, if a node is added to or removed from $\Nl$, then only the $\Delta u$ for the two-hop neighbors needs to be recomputed. 

Simulations suggest that selecting $\n$ in a greedy or a random manner makes little difference for the quality of the computed solution. However, greedy node selection tends to lead to less iterations of the algorithm and is thus better suited for a serial implementation. Random node selection might be preferable for a parallel implementation due to the reduced need for synchronization.

Having computed the set of coordination leaders, there is immediately a vehicle plan for each truck. These plans are jointly optimized as discussed in the following section. 

\section{Joint Vehicle Plan Optimization}
\label{sec:joint_speed-profile_optimization}

In this section, we derive how to jointly optimize the vehicle plans that are selected by the clustering algorithm. We do this by formulating a convex optimization problem with linear constraints for a group consisting of a coordination leader and its coordination followers. 
Hereby, the timing when platoons are assembled and broken apart is adjusted while the locations where this happens is not changed.
Trucks that are not matched to any coordination leader or are not coordination leaders themselves just follow their default plans and are not considered in this section.

\begin{figure}[t]
\begin{center}
 \includegraphics[width=\columnwidth]{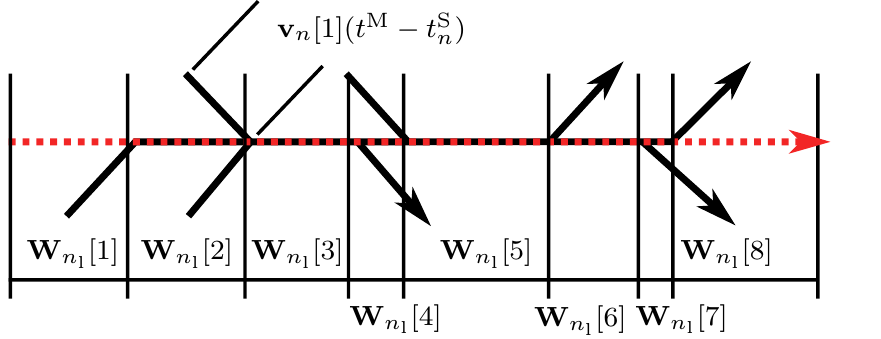}
 \caption{Illustration of how the sequences $\Wconi{n}$ are defined. The red, dotted line represents the route of the coordination leader and the black, solid lines with arrows represent the routes of the coordination followers. The thin lines indicate the distances that the elements of $\Wconi{\nl}$ correspond to.}
\label{fig:coordination_group}
\end{center}
\end{figure}

Consider a coordination leader $\nl \in \Nl$ and its followers $\Nfl{\nl} = \{ n \in \Nc\setminus \Nl : \nl = \arg\max\limits_{i \in \Nl \cap \Nout{n}} \WGcij{n}{i}  \}$. This group of agents is denoted $\Ng = \{\nl\} \cup \Nfl{\nl}$.
We construct an ordered set of time instances $\tcon = (\tconn{1}, \tconn{2}, \dots)$. This set contains the start time and the arrival deadline of the coordination leader, and the merge times and the split times of its followers. 
We divide the distance traveled by the leader from start to destination according to these time instances and get the distances $\Wconin{\nl}{i} = (\tconn{i+1} - \tconn{i})/\vcd$ between these points, where $\vcd$ is the speed of the leader according to its default plan. These are the distances between the points where coordination followers join or leave the platoon.
Similarly, for a coordination follower $n \in \Nfl{\nl}$, we have 
$\Wconi{n} = (\vseqin{n}{1}(\tMi{n} - \tSi{n}),
\Wconin{\nl}{\indMi{n}},  \dots, \Wconin{\nl}{\indSpi{n}},
\vseqin{n}{\vseqNi{n}}(\tAi{n} - \tSpi{n}))$. The variables $\tSi{n}, \tMi{n}, \tSpi{n}, \tAi{n}$ denote the start time, merge time, split time, and arrival time of follower $n$ according to its adapted plan. The first element of $\Wconi{n}$ is the distance along the route from start to the merge point. For the part of the route the follower platoons with the coordination leader, the entries are the same as for the coordination leader. The indices $\indMi{n}, \indSpi{n}$ are defined accordingly. The last element of $\Wconi{n}$ is the distance from the split point to the destination of the follower. Fig.~\ref{fig:coordination_group} illustrates the definition of $\Wconi{n}$.

We introduce sequences $\pconi{n} = (\pconin{n}{1}, \dots, \pconin{n}{|\Wconi{n}|})$ that indicate on which segments of the journey the coordination follower is a platoon follower. If truck $n$ is a platoon follower on the segment that corresponds to $\Wconin{n}{i}$ for some $i$, then $\pconin{n}{i} = 1$. Otherwise we have $\pconin{n}{i} = 0$. 
For the coordination leader $\nl$, we have $\pconi{\nl} = (0,\dots,0)$ and for a coordination follower $n \in \Nfl{\nl}$, we have that $\pconi{n} = (0,1,1,\dots,1,0)$. 

We express the speed and time sequence of truck $n \in \{\nl\} \cup \Nfl{\nl}$ as traversal times $\Tconi{n} = (\Tconin{n}{1}, \dots, \Tconin{n}{|\Wconi{n}|})$ of the segments $\Wconi{n}$. The speed on each such segment remains constant and can be computed as $\vseqin{n}{i} = \Wconin{n}{i}/\Tconin{n}{i}$. The traversal times of the segments in all trucks' routes 
are the optimization variables. Working with traversal times rather than the sequence of speeds $\vseq$ allows us to state the optimization problem with linear constraints. The times when the speed changes $\tseqi{n}$, are computed as $\tseqin{n}{i} = t_{n}^S + \sum\limits_{j = 1}^{i-1} \Tconin{n}{j}$ for $i = 1, \dots, \vseqNi{n}+1$.

With these definitions, we are ready to state the following optimization problem:
\begin{problem}\label{prop:convex_cont_problem}
\begin{subequations}
 \begin{align}
&\min\limits_{ \{\Tconi{n}: n \in \Ng \} } 
\sum\limits_{n \in \Ng}\sum\limits_{i=1}^{\vseqNi{n}} f\left(\frac{\Wconin{n}{i}}{\Tconin{n}{i}},\pconin{n}{i}\right) \Wconin{n}{i} \label{eq:cont_objective}
 \\&\;\;\;\text{s.t.}\nonumber\\
 &\text{for} \; n \in \Ng:\nonumber\\
 &\;\; \frac{\Wconin{n}{i}}{\vmax} \leq \Tconin{n}{i}, \; i \in \{1,\dots,\vseqNi{n}\} \label{eq:cont_opt_v_max_con}\\
 &\;\; \frac{\Wconin{n}{i}}{\vmin} \geq \Tconin{n}{i}, \; i \in \{1,\dots,\vseqNi{n}\} \label{eq:cont_opt_v_min_con}\\
 &\;\; \tSi{n} + \sum\limits_{i=1}^{\vseqNi{n}} \Tconin{n}{i} \leq \tDi{n}\label{eq:cont_opt_t_d_con}\\
 &\text{and for} \; n \in \Nfl{\nl}:\nonumber\\
 &\;\; \tSi{n} + \Tconin{n}{1} = \tSi{\nl} + \sum\limits_{i = 1}^{\indMi{n}-1} \Tconin{\nl}{i} \label{eq:cont_opt_plat_arr_con}\\
 &\;\; \Tconin{n}{1+i} = \Tconin{\nl}{\indMi{n} + i - 1},
 \; i \in \{1,\dots,\indSpi{n} - \indMi{n} + 1\}. \label{eq:cont_opt_plat_trav_con}
\end{align}
\end{subequations}
\end{problem}

The objective function \eqref{eq:cont_objective} equals the combined fuel consumption $\sum\limits_{n \in \Ng} \Fvp{\vfuni{n}}{\pfuni{n}}$ for the assignments $\Ng$ which is part of the sum that defines the combined fuel consumption of all assignment $\Ftot$ defined in \eqref{eq:overall_fuel_consumption}. It is composed of the fuel consumption of the coordination leader and the coordination followers. The coordination leader is considered to travel alone or take the role as the platoon leader throughout its journey. The coordination followers travel alone on the first and the last segment of their journey. They become platoon followers in-between these segments.

There are two sets of constraints. The first set applies to all trucks and ensures that the sequences $\Tconi{n}$ correspond to valid vehicle plans. In particular, the constraints \eqref{eq:cont_opt_v_max_con} and \eqref{eq:cont_opt_v_min_con} express that the trajectories stay within the allowed range of speed. The constraints \eqref{eq:cont_opt_t_d_con} express that all trucks arrive before their deadline. 
The second set of constraints ensures that platooning happens as specified in the original pairwise plans. The constraints \eqref{eq:cont_opt_plat_arr_con} ensure that the coordination leader and each of its followers arrive at the same time at their respective merge point. The constraints \eqref{eq:cont_opt_plat_trav_con} ensure that the speed of the leader and the speed of the follower are the same when they are supposed to platoon. 

When $\fzero$, $\fplat$ are such that $\fzerov{T^{-1}}$ and $\fplatv{T^{-1}}$ are convex in $T$ for $T > 0$, then the objective \eqref{eq:cont_objective} is a sum of convex functions and hence convex. For instance, polynomials with arbitrary constant part and non-negative coefficients fulfill this requirement. Furthermore, all constraints are linear. Thus, Problem~\ref{prop:convex_cont_problem} is a convex optimization problem for which well developed numerical solvers are readily available \cite{convex_opt_book, cvxopt}. 
The optimization is initialized with the pairwise plans.

\section{Simulations}
\label{sec:simulations}

In this section, we evaluate the coordination method outlined in the previous sections with Monte Carlo simulations. We show that coordination of truck platooning can lead to significant reductions in fuel consumption compared to the current situation where trucks do not platoon, as well as compared to spontaneous platooning where trucks only form platoons if they happen to be in the vicinity of another.

\begin{figure}[t]
\begin{center}
 \includegraphics[width=.7\columnwidth]{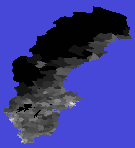}
 \caption{Population density map from which the start and goal locations are sampled. The brighter the pixel, the larger the population density in that area. Areas not belonging to mainland Sweden are shown in blue.}
\label{fig:density_map}
\end{center}
\end{figure}

We generate transport assignments randomly. 
The start and goal locations are sampled within mainland Sweden. The probability of an assignment starting or ending at a particular location is proportional to the population density \cite{population_density}, see Fig.~\ref{fig:density_map}. The resolution is 0.1 degrees in longitude and latitude and the road network node that is closest to the sampled coordinate is chosen. 
We calculate the routes with the Open Source Routing Machine \cite{project_osrm}. 
Assignments for which no route can be found are disregarded.
If the route is longer than 400 kilometers, a 400 kilometers long subsection of the route is randomly selected. This is to take into account that merge points too far from the current position should not be considered for coordination since the uncertainty becomes too large due to traffic, new assignments, and rest periods of the driver. Start locations along the route are considered since we believe that platoon coordination systems will frequently re-plan for assignments that are already en route and suspended for the driver to rest.

The fuel model is an affine approximation around 80\,km/h of the analytical fuel model in \cite{cyberphysicalTransport}. We have for the fuel per distance traveled in kilograms diesel per meter
\begin{align*}
 \fzerov{v} &= 8.4159\cdot 10^{-6} v + 4.8021\cdot 10^{-5} \\
 \fplatv{v} &= 5.0495\cdot 10^{-6} v + 8.5426\cdot 10^{-5}.
\end{align*}
According to this model, the relative reduction in fuel consumption of a platoon follower is 15.9 percent at a speed of 80\,km/h.

We consider a default speed of 80\,km/h and we assume that the speed can be freely chosen between $\vmin = 70$\,km/h and $\vmax = 90$\,km/h throughout the entire journey.
We sample the start time of the assignments uniformly in an interval of 2 hours and compute the arrival deadlines according to the default speed. 

\begin{figure}[t]
\begin{center}
 \includegraphics[width=\columnwidth]{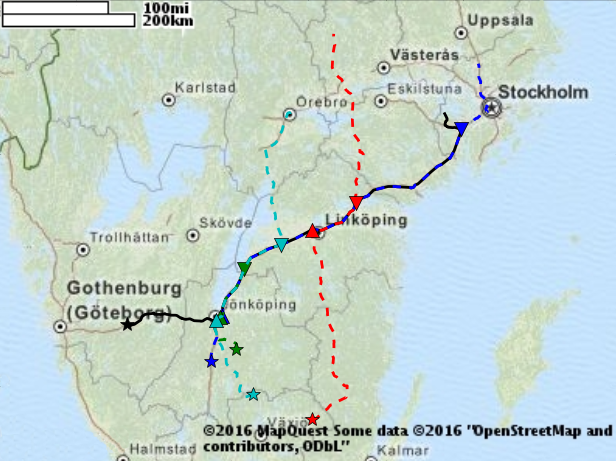}
 \caption{
 The routes of a platoon coordinator with four coordination followers. The route of the coordination leader in shown in black, the routes of the coordination followers are dashed. The beginning of a route is marked with a star. The merge point of a follower is indicated with an upwards-facing triangle and the split point with a downwards-facing triangle.}
\label{fig:platoon_group_map}
\end{center}
\end{figure}

The pairwise plans are such that trucks platoon as long as possible. Once a coordination follower splits up from the coordination leader, it drives fast enough to arrive in time at its destination and at least at default speed. The split points are such that arriving in time is feasible. Thus, trucks are guaranteed to meet their deadlines and the initial value for the joint vehicle plan optimization fulfills the constraints. 
Fig.~\ref{fig:platoon_group_map} shows an example of the routes of a coordination leader and its coordination followers and where the coordination followers join and leave the platoon. 

We compare our proposed platoon coordinator to fuel savings that arise from spontaneous platooning, i.e., that trucks happen to get into each others vicinity and then spontaneously form platoons. To this end, we collect all the link arrival times according to the default plans for each link in the scenario. We sort these times and collect them in ascending order in groups of at most one minute difference in their edge arrival time. We assume that each of these groups forms a platoon driving at default speed and that the default trajectory is not altered by the platooning. This is a generous estimate since it neglects any kind of coordination effort which would be present for time gaps up to one minute.

In order to assess the the quality of the solution computed by the clustering algorithm, we establish an upper bound on the solution of Problem~\ref{prob:clustering}.
This upper bound is based on the intuition to assign every truck its best coordination leader and ignore that coordination leaders do not contribute to the objective.
We have that
\begin{equation}\label{eq:upper_bound}
\begin{split}
 \fce(\Nl) 
 &= 
 \sum\limits_{i \in \Nc \setminus \Nl} \max_{j \in \Nout{i} \cap \Nl}  \WGcij{i}{j} 
 \\
 &\leq
 \sum\limits_{i \in \Nc \setminus \Nl} \max_{j \in \Nout{i}} \WGcij{i}{j} 
 \\
 &\leq 
 \sum\limits_{i \in \Nc} \max_{j \in \Nout{i}}  \WGcij{i}{j},
\end{split}
\end{equation}
where the second inequality holds since $\WGcij{i}{j} > 0$ for all $(i,j) \in \set{E}$. 

This bound can only be tight when there is an optimal solution where no coordination leader has an out-neighbor. Otherwise the coordination leaders cannot contribute to the sum. Nevertheless, the bound helps us assess how far a heuristic solution can be away from the optimum.

We implemented platoon coordination in Python and used CVXOPT \cite{cvxopt} for convex optimization. The execution of the clustering algorithm takes less than a second for 2000 transport assignments. Even faster computation times could be achieved by optimizing the implementation. 

Each simulation consists of the following steps:
\begin{enumerate}
 \item Random generation of transport assignments
 \item Computation of routes and default plans
 \item Computation of the coordination graph
 \item Computation of coordination leaders according to Section~\ref{sec:clustering_alg}
 \item Joint vehicle plan optimization according to Section~\ref{sec:joint_speed-profile_optimization}
\end{enumerate}

We evaluate how different numbers of assignments affect the amount of platooning and the fuel savings relative to the default plans.
For comparison we compute the fuel savings of spontaneous platooning. We run the clustering algorithm with greedy and random node selection and compute the upper bound of the objective function $\fce$. The results are averaged over 150 simulation runs.

\begin{figure}[t]
\begin{center}
 \includegraphics[width=.6\columnwidth]{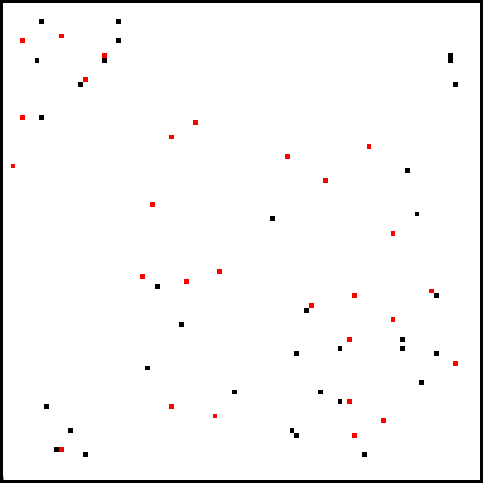}
 \caption{
 This plots visualizes the adjacency matrix of a coordination graphs with 100 assignments. Nonzero entries are indicated with a black or a red dot, each corresponding to an edge in the coordination graph. Edges whose corresponding plans are selected by the clustering algorithm correspond to the red dots.
 }
\label{fig:adjecency}
\end{center}
\end{figure}

Fig.~\ref{fig:adjecency} visualizes an example coordination graph. In addition it shows which assignments are selected in step 4). We can see that only a small fraction of assignment pairs can save fuel by forming a platoon. As the number of assignments grows, more opportunities are available for each assignment which can translate into larger fuel savings \cite{distributed_platooning}.

\begin{figure}[t]
\begin{center}
 \includegraphics[width=\columnwidth]{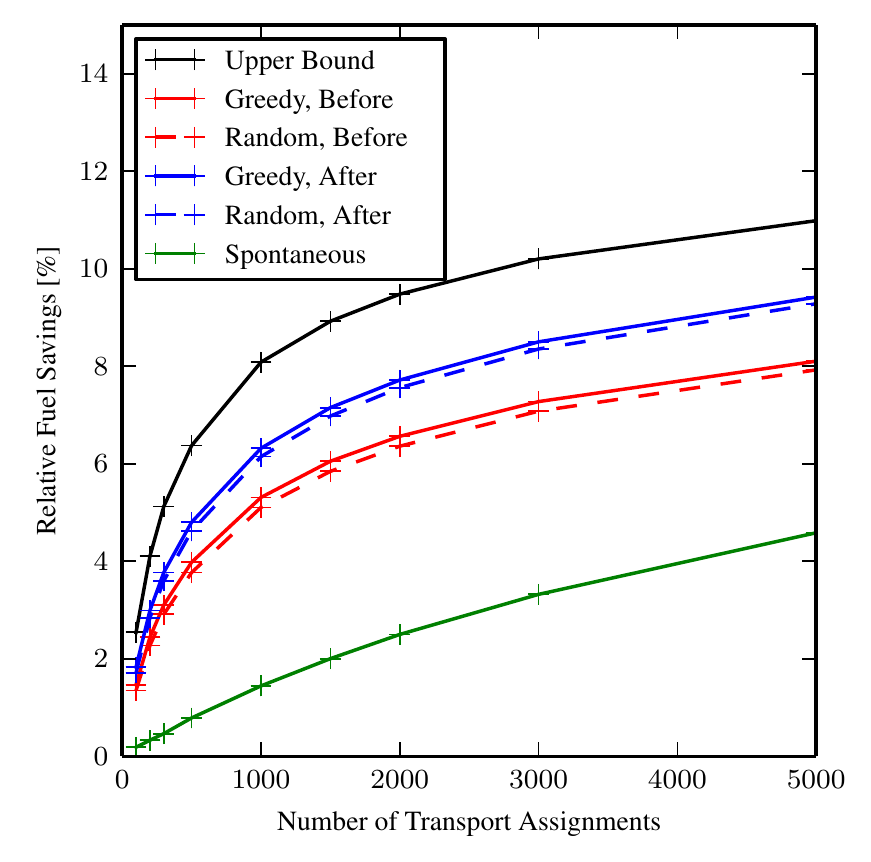}
 \caption{The relative fuel savings due to platooning compared to the default plans with varying numbers of assignments. ``Greedy'' indicates that greedy node selection was used in the clustering algorithm, whereas ``Random'' indicates random node selection. The keywords ``Before''/``After'' refer to the relative fuel savings before/after the joint optimization of the vehicle plans. ``Spontaneous'' are the relative fuel savings based on the estimate of fuel savings due to spontaneous platooning. ``Upper Bound'' refers to the upper bound the $\fce$ as stated in \eqref{eq:upper_bound}.}
\label{fig:relative_fuel_savings}
\end{center}
\end{figure}

Fig.~\ref{fig:relative_fuel_savings} shows the effect on the fuel savings when the numbers of transport assignments that are coordinated is varied. It is possible to make a number of observations based on these data. First of all, the fuel savings increase rapidly with the number of transport assignments when the absolute number of assignments is small. As more and more assignments are added, this trend stagnates and the relative fuel savings increase only slowly. Ideally this should approach asymptotically the maximum fuel savings of 15.9\,\% as the number of transport assignments goes to infinity, since then virtually every truck is a platoon follower for its entire journey. There is only a small difference between greedy and random node selection, however, with the greedy node selection outperforming the random node selection consistently. For a parallel or even a distributed implementation of the clustering algorithm, random node selection would be preferable due to the reduced need for synchronization whereas greedy node selection is faster in a centralized setting. Furthermore, the results after selecting the coordination leaders and before the joint convex optimization are less than the upper bound but only about 30\,\% worse. Since the upper bound is not tight, this indicates that the clustering algorithm performs well. We can see a clear improvement in the fuel savings by the joint optimization of the vehicle plans. Spontaneous platooning gives fuel savings that are less than half of what can be achieved by coordination. Also bear in mind that this is a generous estimate of fuel savings by spontaneous platooning so that the real difference would probably be even larger. 

We conclude that coordinated platooning can yield significant fuel savings and that coordination is crucial leveraging these savings. For 2000 transport assignments starting over the course of two hours, we get 7.6\,\% reduction in fuel consumption. A number of 2000 trucks starting in that time interval on an area like Sweden is a realistic number. The total distance traveled in the simulated scenario is in the same order of magnitude as the total distance traveled by domestic road freight transport in Sweden within two hours, assuming that traffic volume is equally spread over the year  \cite{lastbilstrafik}. The density of the road freight traffic that was simulated is only a fraction of the total road freight traffic in countries with high population density. The small fuel savings for the platoon leaders, which have been neglected, would further increase the platooning benefit.

\begin{figure}[t]
\begin{center}
 \includegraphics[width=\columnwidth]{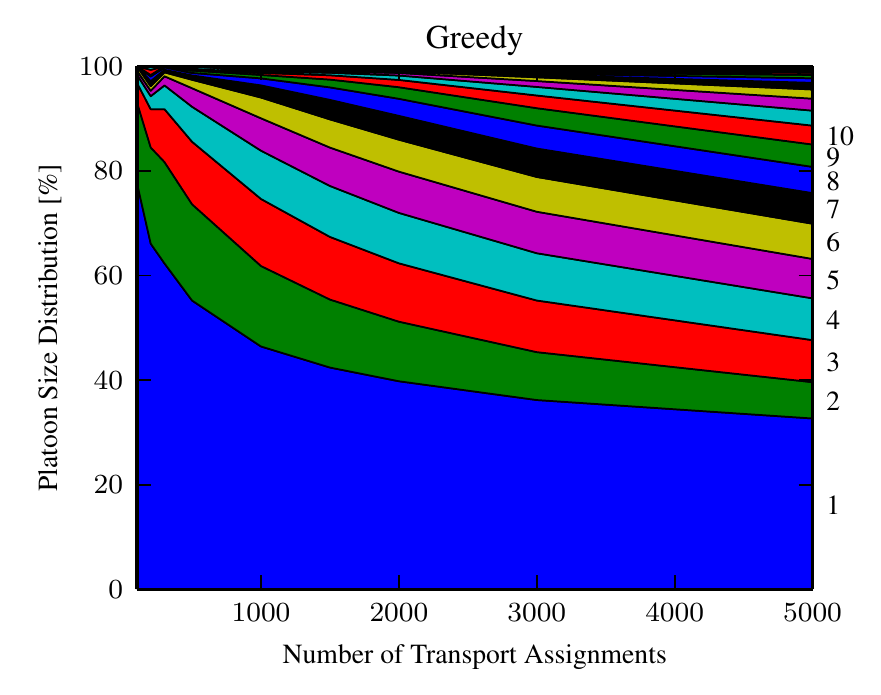}
 \includegraphics[width=\columnwidth]{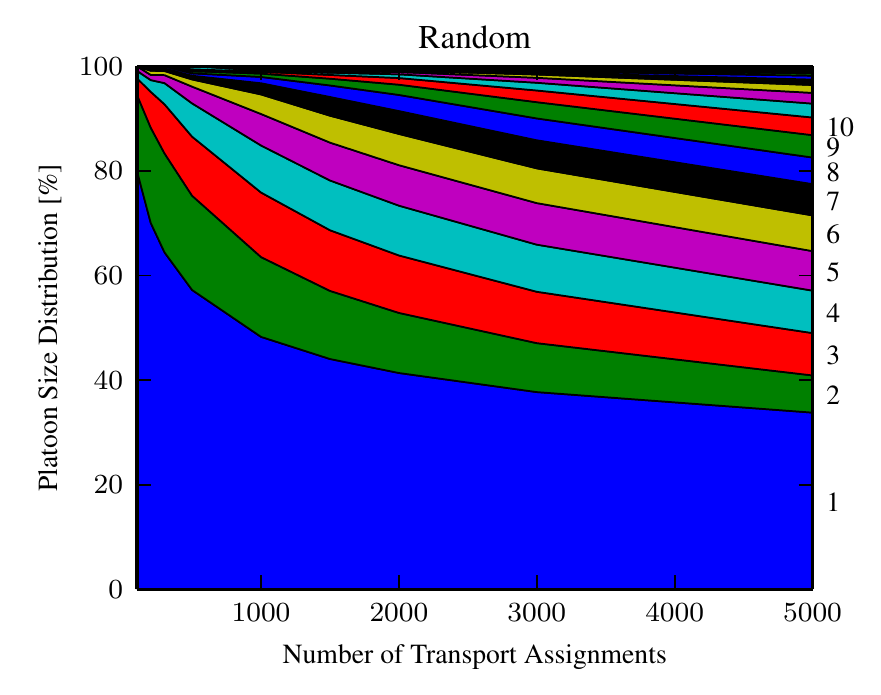}
 \caption{This figure shows the distribution of platoon sizes per distance traveled over the number of assignments in percent. The upper plot shows the results of greedy node selection whereas the lower plot shows those of random node selection in the clustering algorithm. To the right, the size of platoon is indicated for a platoon size up to ten. So, when the difference between the first and the second boundary from below is for instance at 20\,\%, it means that 20\,\% of the distance was traveled as member of a platoon of size 2.}
\label{fig:platoon_sizes}
\end{center}
\end{figure}

Fig.~\ref{fig:platoon_sizes} shows how the distribution of platoon sizes changes with the number of transport assignments. We can see that the larger the number of transport assignments, the more distance is traveled in large platoons. For 2000 assignments over half the distance traveled is in a platoon. Most of the distance is traveled in platoons with ten or less vehicles. This is promising since large platoons might be difficult to control and thus the platoon coordinator would have to prevent planning for larger platoons. Since these large platoons only account for a small fraction of the distance traveled, this would not have too large an impact on the total fuel savings. The largest platoon formed has 28 vehicles. A noticeable effect occurs at a number of 200 transport assignments when more distance is traveled in relatively large platoons compared to the distribution with a number of 300 transport assignments. It seems that some kind of phase transition occurs at these points, where enough assignments are in the system to go from one coordination leader with many followers to having several coordination leaders that are better suited for their followers. To understand this phenomenon is subject of future work.

The simulations show that computing plans for a large number of vehicles to form platoons is feasible with the methods outlined in this paper. It motivates that real-time platoon coordination enables significant reductions in fuel consumption and might be the key to leveraging the full potential of truck platooning.

\section{Conclusion}

A centralized truck platoon coordinator was proposed. The system provides trucks with vehicle plans that lead to reduced fuel consumption by making use of platooning. As time evolves, plans are updated to account for deviations and new assignments. In order to handle the complexity of such coordination, the problem was formulated in a way that for each truck a number of plans adapted to the default plans of other vehicles are computed. Each adapted plan involves platooning for some distance as platoon follower and thus saving fuel. It was derived how these plans should be systematically combined in order to maximize the total fuel savings. The NP-hardness of this problem motivates the proposed heuristic solution method. Furthermore, we derived how to jointly optimize the vehicle plans resulting from the combination of default plans and adapted plans. The effectiveness of the method was demonstrated in a realistic simulation study. The simulations motivate that such systems should be deployed once trucks with the ability to platoon are commercially available. 

There are various directions of future work. One direction is to understand how the transport assignments and the road network relate to fuel savings achieved by this method. Furthermore, we want to study a receding horizon implementation of the platoon coordinator under the presence of disturbances. Another direction is to study the system in a setting where more practical details such as speed limits, traffic, driver rest times, different vehicle types etc. are taken into account. Finally, similar coordination strategies might be relevant for other types of multi-agent systems.  This work was carried out in the scope of the COMPANION EU project~\cite{companion_overview} and the proposed platoon coordinator is being implemented in a demonstrator featuring simulated and real trucks. 

\appendix[Proof of Proposition~\ref{prop:NP}]

 We show the result by reduction of the optimization version of the set covering problem to Problem~\ref{prob:clustering}. The optimization version of the set covering problem is well known to be NP-hard. Reduction to a known hard problem is a common proof technique for this kind of result \cite{intro_to_algorithms}.
 We do this by constructing a coordination graph $\Gc$ for which there is a one-to-one correspondence between coordination leaders and selected sets for the cover. Then we show that the minimum number of leaders that corresponds to a set cover gives the maximum value for $\fce$.
 
 Consider the following set covering problem. We have a finite set $\U$. Furthermore, let $\FS$ be a family of subsets of $\U$ with $\bigcup\limits_{\sps \in \FS} \sps = \U$. The problem is to find the smallest number of subsets in $\FS$ whose union is $\U$. 
 
\begin{figure}[t]
\begin{center}
 \includegraphics[width=\columnwidth]{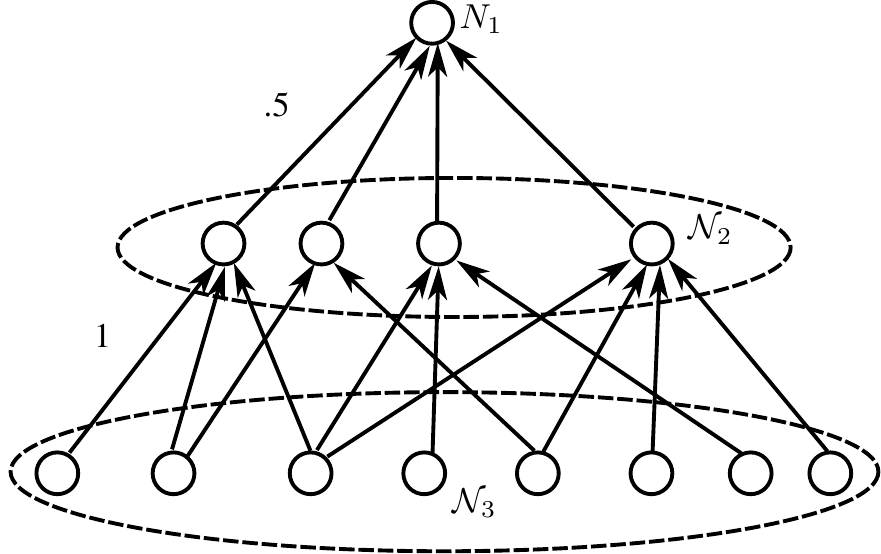}
 \caption{Illustration of the graph used to prove that Problem~\ref{prob:clustering} is NP-hard.}
\label{fig:graph.pdf}
\end{center}
\end{figure}

 We construct the coordination graph as the one shown in Fig.~\ref{fig:graph.pdf}.
 We introduce a node for each element in $\U$. We denote the set of these nodes with $\NB$ and let $\mB: \U \rightarrow \NB$ be a bijective mapping from the elements in $\U$ to the nodes in $\NB$. We introduce a node for each element in $\FS$. 
 We denote the set of these nodes with $\NA$ and let $\mA: \FS \rightarrow \NA$ be a bijective mapping from elements in $\FS$ to nodes in $\NA$. Consider a node $n_2 \in \NA$ that corresponds to the element $\SpS \in \FS$.
 The in-neighbors of $n_2$ are $\Nin{n_2} = \{\mB(\sps): \sps \in \mA^{-1}(n_2)\}$.  
 The weight of the corresponding edges is $1$. We introduce an additional node $\NO$. There is an edge from each node in $\NA$ to $\NO$ with weight $0.5$. Clearly, this reduction is linear in the size of the input $\U, \FS$. 

 Since $\NO$ has no out-neighbors, its membership in $\Nl$ can only increase $\fce(\Nl)$. Since all nodes in $\NB$ have no in-neighbors, adding a node in $\NB$ to $\Nl$ can only decrease $\fce(\Nl)$. Thus, the problem of finding the optimal $\Nl$ reduces to finding which nodes in $\NA$ belong to $\Nl$.
 In the optimal solution, each node in $\NB$ has at least one out-neighbor in $\Nl$. Otherwise we could add any out-neighbor of that node to $\Nl$ and increase $\fce(\Nl)$ by at least $0.5$. 
 Therefore, $\{\mA^{-1}(n): n \in \Nl \cap \NA\}$ is a set cover of $\U$. Otherwise there would be $u \in \U$ such that there is no $\SpS \in \{\mA^{-1}(n): n \in \Nl \cap \NA\}$ with $u \in \SpS$. If such a $u$ existed, $\mB(u)$ would be a node with no out-neighbor in $\Nl \cap \NA$. 
 Furthermore let $\bar{\FS} \subseteq \FS$ be a set cover of $\U$. Then $\{\mA(\sps): \sps \in \bar{\FS} \}$ has the property that $\bigcup\limits_{n \in \{\mA(\sps): \sps \in \bar{\FS}\} } \Nin{n} = \NB$, so any set cover has the property that all nodes in $\NB$ have at least one out-neighbor in $\Nl$. 
 Each node in $\NA$ contributes with $0.5$ to the objective if it is not in $\Nl$. Therefore, the optimal $\Nl$ contains a minimum number of nodes from $\NA$ such that every node in $\NB$ has at least one out-neighbor in $\Nl \cap \NA$. Since any $\Nl \cap \NA$ that fulfills this property maps to a set cover $\bar{\FS}$ and vice versa, and since $|\Nl \cap \NA| = |\bar{\FS}|$, we have that $\bar{\FS}$ is the solution to the set covering problem. Thus the NP-hard set-covering problem can be reduced to Problem~\ref{prob:clustering}, which shows that Problem~\ref{prob:clustering} is NP-hard.

\section*{Acknowledgments}

This work was supported by the COMPANION EU project, the Knut and Alice Wallenberg Foundation, the Swedish Strategic Research Foundation, and the Swedish Research Council.

\ifCLASSOPTIONcaptionsoff
  \newpage
\fi

% \IEEEtriggercmd{\enlargethispage{2mm}}


% \bibliographystyle{IEEEtran}
% % argument is your BibTeX string definitions and bibliography database(s)
% \bibliography{IEEEabrv,citations_reduced}

\begin{thebibliography}{10}
\providecommand{\url}[1]{#1}
\csname url@samestyle\endcsname
\providecommand{\newblock}{\relax}
\providecommand{\bibinfo}[2]{#2}
\providecommand{\BIBentrySTDinterwordspacing}{\spaceskip=0pt\relax}
\providecommand{\BIBentryALTinterwordstretchfactor}{4}
\providecommand{\BIBentryALTinterwordspacing}{\spaceskip=\fontdimen2\font plus
\BIBentryALTinterwordstretchfactor\fontdimen3\font minus
  \fontdimen4\font\relax}
\providecommand{\BIBforeignlanguage}[2]{{%
\expandafter\ifx\csname l@#1\endcsname\relax
\typeout{** WARNING: IEEEtran.bst: No hyphenation pattern has been}%
\typeout{** loaded for the language `#1'. Using the pattern for}%
\typeout{** the default language instead.}%
\else
\language=\csname l@#1\endcsname
\fi
#2}}
\providecommand{\BIBdecl}{\relax}
\BIBdecl

\bibitem{path_overview_conference}
R.~Horowitz and P.~Varaiya, ``Control design of an automated highway system,''
  \emph{Proc. {IEEE}}, vol.~88, no.~7, pp. 913--925, Jul. 2000.

\bibitem{Bonnet2000}
C.~Bonnet and H.~Fritz, ``Fuel consumption reduction in a platoon: Experimental
  results with two electronically coupled trucks at close spacing,''
  \emph{{SAE} Tech. Paper 2000-01-3056}, 2000.

\bibitem{lammert_fuel_consumption}
M.~P. Lammert \emph{et~al.}, ``Effect of platooning on fuel consumption of
  class 8 vehicles over a range of speeds, following distances, and mass,''
  \emph{SAE Int. J. Commer. Veh.}, vol.~7, pp. 626--639, Sep. 2014.

\bibitem{ITS_overview}
S.~Tsugawa, ``An overview on an automated truck platoon within the {Energy ITS
  Project},'' \emph{Advances in Automotive Control}, vol.~7, pp. 41--46, 2013.

\bibitem{FreightPlanningSurvey}
T.~G. Crainic and G.~Laporte, ``Planning models for freight transportation,''
  \emph{European J. Operational Research}, vol.~97, no.~3, pp. 409--438, 1997.

\bibitem{eco_routing_journal}
K.~Boriboonsomsin \emph{et~al.}, ``Eco-routing navigation system based on
  multisource historical and real-time traffic information,'' \emph{{IEEE}
  Trans. Intell. Transp. Syst.}, vol.~13, no.~4, pp. 1694--1704, Dec. 2012.

\bibitem{road_traffic_control}
M.~Papageorgiou \emph{et~al.}, ``Review of road traffic control strategies,''
  \emph{Proc. {IEEE}}, vol.~91, no.~12, pp. 2043--2067, Dec. 2003.

\bibitem{Larsson_Platoon_Complexity}
E.~Larsson \emph{et~al.}, ``The vehicle platooning problem: Computational
  complexity and heuristics,'' \emph{Transportation Research Part {C}: Emerging
  Technologies}, vol.~60, pp. 258--277, 2015.

\bibitem{kuoyun_catchup}
K.-Y. Liang \emph{et~al.}, ``When is it fuel efficient for a heavy duty vehicle
  to catch up with a platoon?'' in \emph{7th IFAC Symp. Advances in Automotive
  Control}, 2013.

\bibitem{jeff_kuo_yun_distributed_controller}
J.~Larson \emph{et~al.}, ``A distributed framework for coordinated heavy-duty
  vehicle platooning,'' \emph{{IEEE} Trans. Intell. Transp. Syst.}, vol.~16,
  no.~1, pp. 419--429, Feb. 2015.

\bibitem{datamining_platooning}
P.~Meisen \emph{et~al.}, ``A data-mining technique for the planning and
  organization of truck platoons,'' in \emph{Int. Conf. Heavy Vehicles, Heavy
  Vehicle Transport Technology}, 2008, pp. 389--402.

\bibitem{ACCpaper}
S.~van~de Hoef \emph{et~al.}, ``Fuel-optimal coordination of truck platooning
  based on shortest paths,'' in \emph{American Control Conf.}, 2015.

\bibitem{ITSCpaper}
------, ``Coordinating truck platooning by clustering pairwise fuel-optimal
  plans,'' in \emph{18th IEEE Int. Conf. Intelligent Transportation Syst.},
  2015.

\bibitem{pam_book}
L.~Kaufman and P.~J. Rousseeuw, \emph{Finding Groups in Data: An introduction
  to Cluster Analysis}.\hskip 1em plus 0.5em minus 0.4em\relax John Wiley \&
  Sons, Inc., 2008.

\bibitem{clustering_book1}
A.~K. Jain and R.~C. Dubes, \emph{Algorithms for Clustering Data}.\hskip 1em
  plus 0.5em minus 0.4em\relax Upper Saddle River, NJ, USA: Prentice-Hall,
  Inc., 1988.

\bibitem{clustering_overview_paper}
A.~K. Jain, ``Data clustering: 50 years beyond {K}-means,'' \emph{Pattern
  Recognition Lett.}, vol.~31, no.~8, pp. 651--666, 2010.

\bibitem{blondelcommunity}
V.~Blondel \emph{et~al.}, ``Fast unfolding of communities in large networks,''
  \emph{J. Stat. Mech.}, P10008, 2008.

\bibitem{community_detection_survey}
S.~Harenberg \emph{et~al.}, ``Community detection in large-scale networks: a
  survey and empirical evaluation,'' \emph{Wiley Interdisciplinary Reviews:
  Computational Stat.}, vol.~6, no.~6, pp. 426--439, 2014.

\bibitem{community_detection_survey2}
S.~Fortunato, ``Community detection in graphs,'' \emph{Physics Rep.}, vol. 486,
  no. 3–5, pp. 75--174, 2010.

\bibitem{bully_alg}
H.~Garcia-Molina, ``Elections in a distributed computing system,'' \emph{{IEEE}
  Trans. Comput.}, vol. C-31, no.~1, pp. 48--59, 1982.

\bibitem{electing_good_leaders}
S.~Singh and J.~Kurose, ``Electing ``good'' leaders,'' \emph{J. Parallel and
  Distributed Computing}, vol.~21, no.~2, pp. 184--201, 1994.

\bibitem{fuel_model_review}
E.~Demir \emph{et~al.}, ``A review of recent research on green road freight
  transportation,'' \emph{European J. Operational Research}, vol. 237, no.~3,
  pp. 775--793, 2014.

\bibitem{routing_overview}
P.~Sanders and D.~Schultes, ``Engineering fast route planning algorithms,'' in
  \emph{Proc. Experimental Algorithms: 6th International Workshop}, Rome,
  Italy, Jun. 2007, pp. 23--36.

\bibitem{intro_to_algorithms}
T.~H. Cormen \emph{et~al.}, \emph{Introduction to Algorithms}, 3rd~ed.\hskip
  1em plus 0.5em minus 0.4em\relax {MIT Press}, 2009.

\bibitem{convex_opt_book}
S.~Boyd and L.~Vandenberghe, \emph{Convex Optimization}.\hskip 1em plus 0.5em
  minus 0.4em\relax New York, NY: Cambridge University Press, 2004.

\bibitem{cvxopt}
\BIBentryALTinterwordspacing
M.~S. Andersen \emph{et~al.} (2013) \textit{{CVXOPT}: A {P}ython package for
  convex optimization}. [Online]. Available: \url{cvxopt.org}
\BIBentrySTDinterwordspacing

\bibitem{population_density}
{Socioeconomic Data and Application Center}, ``Population density grid, v3,
  2000,'' 2015.

\bibitem{project_osrm}
D.~Luxen and C.~Vetter, ``Real-time routing with openstreetmap data,'' in
  \emph{Proc. 19th ACM SIGSPATIAL Int. Conf. Advances in Geographic Inform.
  Syst.}, ser. GIS '11, New York, NY, USA, 2011, pp. 513--516.

\bibitem{cyberphysicalTransport}
B.~Besselink \emph{et~al.}, ``Cyber-physical control of road freight
  transport,'' \emph{Proc. {IEEE}}, 2016.

\bibitem{distributed_platooning}
J.~Larson \emph{et~al.}, ``A distributed framework for coordinated heavy-duty
  vehicle platooning,'' \emph{{IEEE} Trans. Intell. Transp. Syst.}, vol.~16,
  no.~1, pp. 419--429, Feb. 2015.

\bibitem{lastbilstrafik}
``{Swedish} national and international road goods transport 2014,'' Transport
  Analysis Stockholm, Tech. Rep., 2015.

\bibitem{companion_overview}
S.~Eilers \emph{et~al.}, ``{COMPANION -- Towards Co-operative Platoon
  Management of Heavy-Duty Vehicles},'' in \emph{18th IEEE Int. Conf.
  Intelligent Transportation Syst.}, 2015.

\end{thebibliography}

% \vspace{20mm}
\IEEEtriggeratref{7}
% \IEEEtriggercmd{\enlargethispage{-80mm}}

\newpage

\end{document}